\documentclass[12pt]{article}
\usepackage{epsfig}
\newcommand{\be}{\begin{equation}}
\newcommand{\ee}{\end{equation}}
\newcommand{\bea}{\begin{eqnarray}}
\newcommand{\eea}{\end{eqnarray}}
\newcommand{\norsl}{\normalsize\sl}
\newcommand{\norsc}{\normalsize\sc}
\newcommand{\bi}{\begin{itemize}}
\newcommand{\ei}{\end{itemize}}
\makeatletter
         
          \@addtoreset{equation}{section}
\makeatother

\begin{document}

\begin{titlepage}

\title{ Virtual Photon Structure Functions \\
and Positivity Constraints
}

\author{
\norsc  Ken SASAKI~$^a$\thanks{e-mail address: sasaki@cnb.phys.ynu.ac.jp},~
           Jacques SOFFER~$^b$\thanks{e-mail address: soffer@cpt.univ-mrs.fr}~
and       Tsuneo UEMATSU~$^c$\thanks{e-mail address:
uematsu@phys.h.kyoto-u.ac.jp}
\\
\norsl  $^a$ Dept. of Physics,  Faculty of Engineering, Yokohama National
University \\
\norsl  Yokohama 240-8501, JAPAN \\
\norsl $^b$ Centre de Physique Th{\'e}orique, CNRS, Luminy Case 907,
\\
\norsl        F-13288 Marseille Cedex 9, FRANCE \\
\norsl $^c$ Dept. of Fundamental Sciences, FIHS, Kyoto University \\
\norsl     Kyoto 606-8501, JAPAN \\
}

\date{}
\maketitle

\begin{abstract}
{\normalsize
We study the three positivity constraints on the eight virtual photon structure 
functions, derived from the Cauchy-Schwarz inequality and which are
hence, model-independent. The photon structure functions obtained from the 
simple parton model show quite different behaviors in a massive or a 
massless quark case, but they satisfy, in both cases, the three positivity 
constraints. We then discuss an inequality which holds among the unpolarized and
polarized photon structure functions $F_1^\gamma$, $g_1^\gamma$ and 
$W_{TT}^\tau$, in the kinematic region $\Lambda^2\ll P^2 \ll Q^2$, where 
$-Q^2 (-P^2)$ is the mass squared of the probe (target) photon, and we
examine whether this inequality is satisfied by the perturbative QCD results.
}
\end{abstract}

\begin{picture}(5,2)(-290,-580)
\put(2.3,-65){YNU-HEPTh-02-102}
\put(2.3,-80){CPT-2002/PE.4346}
\put(2.3,-95){KUCP-207}
\put(2.3,-110){May 2002}
\end{picture}

\thispagestyle{empty}
\end{titlepage}
\setcounter{page}{1}
\baselineskip 18pt

\section{Introduction}
The investigation of the photon structure is an active field of
research both theoretically and  experimentally~\cite{RN}. Structure
functions of unpolarized real and virtual photons, $F_2^\gamma$ and
$F_{eff}^{\gamma^*}$, have been measured through the two-photon processes 
in $e^+e^-$
collisions as well as the resolved  photon processes in the $ep$ collider.
 From these data the unpolarized parton
distributions in the photon were extracted in the framework of perturbative 
QCD (pQCD)
\cite{MK}. On the other hand, there has been growing interest in the study 
of polarized
photon structure functions \cite{Barber,SVZ}. Especially the first moment of
the spin-dependent  structure function $g_1^\gamma$ has attracted much
attention in connection with its relevance to the QED and QCD axial anomaly.
The next-to-leading order (NLO) QCD analysis
of $g_1^\gamma$ was performed \cite{SV,SU,GRS} and, recently, the second
spin-dependent  structure function $g_2^\gamma$ of  virtual photon has been 
studied
in conjunction with the twist-3 contribution~\cite{BSU}.
For the real photon structure functions, $g_1^\gamma$ and $F_1^\gamma$,
 there exists a positivity bound, $|g_1^\gamma|\leq 
F_1^\gamma$.
This bound has been analyzed recently in detail in Ref.\cite{GRS}.

Now we note that
there appear, in total, eight structure functions in the case of virtual photon
target \cite{BCG,BGMS,BM,CW}, most of which have not been measured yet and,
therefore, are unknown. In such a situation, positivity bounds would play
an important role in constraining these unknown structure functions.
It is well known in deep inelastic scattering off nucleon, that
various bounds have been obtained for the spin-dependent observables 
and parton distributions in a nucleon by means of positivity conditions
\cite{JS1}. In our previous paper \cite{SSU} we have derived
three positivity bounds, among the eight virtual photon
structure functions, which hold model-independently. The number of positivity
bounds reduces to one in the real photon case, and we have checked that 
this remaining bound is indeed satisfied by the structure functions 
obtained in the simple parton model (PM). We also presented a positivity 
bound for the quark distributions relevant for the spin-dependent 
semi-inclusive process in the two-photon reaction.

In this paper we examine the three positivity constraints on the virtual
 photon structure functions. By evaluating the box (a quark-loop)
diagrams, we first obtain the eight virtual photon structure functions in 
the PM and check if they satisfy the positivity constraints or not.
We then discuss an inequality which holds among the unpolarized and polarized
structure functions, $F_1^\gamma$,
$g_1^\gamma$ and $W_{TT}^\tau$, and we examine the pQCD results.

In the next section we discuss the eight virtual photon structure functions 
which were introduced by Budnev et al.~\cite{BCG} to describe the 
absorptive part
of the virtual photon-photon forward scattering. The positivity 
constraints, which were
derived in our previous paper~\cite{SSU} for the eight independent 
$s$-channel helicity
amplitudes, are rewritten in terms of these structure functions.
In Sec. 3 we calculate these eight structure functions in the PM.
We find that there exists a clear difference, both in the $x$-dependence
and in the magnitude, between the massive and the massless quark cases
for the PM predictions. But for both cases, it turns out that the
three positivity constraints are indeed satisfied for all the allowed $x$ region.
 In Sec. 4 we study an inequality which holds among $F_1^\gamma$, $g_1^\gamma$ and 
$W_{TT}^\tau$,
in the kinematic region $\Lambda^2\ll P^2 \ll Q^2$, where $-Q^2 (-P^2)$
is the mass squared of the probe (target) photon. Since the NLO QCD results 
for $F_1^\gamma$ and $g_1^\gamma$ and the leading order (LO) result for $W_{TT}^\tau$ are 
already known, 
we will examine whether these pQCD results are consistent with this 
inequality. The last section is devoted to the conclusion.

\section{Photon structure functions and positivity constraints}

We consider the virtual photon-photon forward scattering amplitude for
$\gamma(q)+\gamma(p)\rightarrow \gamma(q)+\gamma(p)$  illustrated in Fig.1,
\be
T_{\mu\nu\rho\tau}(p,q)=i\int d^4 x d^4 y d^4 z e^{iq\cdot x}e^{ip\cdot (y-z)}
\langle 0|T(J_\mu(x) J_\nu(0) J_\rho(y) J_\tau(z))|0\rangle~,
\ee
where $J$ is the electromagnetic current,  $q$ and $p$ are the four-momenta of the
probe and
target photon, respectively. The $s$-channel helicity amplitudes are related
to
its absorptive part as follows:
\be
W(ab\vert a'b')=\epsilon^*_\mu(a)\epsilon^*_\rho(b)
W^{\mu\nu\rho\tau}\epsilon_\nu(a')\epsilon_\tau(b')~,
\ee
where
\be
W_{\mu\nu\rho\tau}(p,q)=\frac{1}{\pi}{\rm Im}T_{\mu\nu\rho\tau}(p,q)~,
\ee
and $\epsilon_\mu (a)$ represents the photon polarization
vector with helicity $a$, and $a =0, \pm1$. Similarly for
the other polarization vectors and we have  $a', b, b'=0, \pm1$.
Due to angular momentum conservation, parity conservation and time
reversal
invariance \cite{BLS}, we have in total eight independent $s$-channel helicity
amplitudes,
which we may take as
\bea
&&W(1,1\vert 1,1),\ \
W(1,-1\vert 1,-1),\ \  W(1,0\vert 1,0),\ \  W(0,1\vert 0,1),\ \  W(0,0\vert
0,0),
\nonumber  \\
&&W(1,1\vert -1,-1),\ \  W(1,1\vert 0,0),\ \  W(1,0\vert 0,-1).
\eea
The first five amplitudes are helicity-nonflip  and the last three are
helicity-flip. It is noted that the $s$-channel helicity-nonflip amplitudes
are semi-positive, but not the helicity-flip ones.

In our previous work~\cite{SSU}, we have applied the Cauchy-Schwarz
inequality~\cite{JS,JT}  to the above photon helicity amplitudes and have
derived a
positivity bound:
\be
\Bigl|W(a,b\vert a',b')  \Bigr|\leq \sqrt{W(a,b\vert a,b)
W(a',b'\vert a',b')}~.
\ee
Writing down explicitly, we obtain
the following three positivity constraints:
\bea
\Bigl|W(1,1\vert -1,-1)  \Bigr|&\leq& W(1,1\vert 1,1)~,  \label{CS1}\\
\Bigl|W(1,1\vert 0,0)  \Bigr|&\leq& \sqrt{W(1,1\vert 1,1)
W(0,0\vert 0,0)}~,\label{CS2}\\
\Bigl|W(1,0\vert 0,-1)  \Bigr|&\leq& \sqrt{W(1,0\vert 1,0)W(0,1\vert 0,1)}~.
\label{CS3}
\eea

The photon-photon scattering phenomenology is often discussed in terms of
the photon structure functions instead of the $s$-channel helicity amplitudes.
Budnev, Chernyak and Ginzburg [BCG]
\cite{BCG}  introduced the following eight independent structure functions, in
terms of
which  the absorptive part of virtual photon-photon forward scattering,
$W^{\mu\nu\rho\tau}$, is written as,
\bea
W^{\mu\nu\rho\tau}(p,q)&=&({P_{TT}})^{\mu\nu\rho\tau}W_{TT}+
                     ({P^a_{TT}})^{\mu\nu\rho\tau}W^a_{TT}+
                    ({P^{\tau}_{TT}})^{\mu\nu\rho\tau}W^{\tau}_{TT} \nonumber
\\
                 &+&({P_{ST}})^{\mu\nu\rho\tau}W_{ST}+
                      ({P_{TS}})^{\mu\nu\rho\tau}W_{TS}+
                           ({P_{SS}})^{\mu\nu\rho\tau}W_{SS} \nonumber  \\
     &-& ({P^{\tau}_{TS}})^{\mu\nu\rho\tau}W^{\tau}_{TS}
         - ({P^{\tau a}_{TS}})^{\mu\nu\rho\tau}W^{\tau a}_{TS}~,
\label{Wmunurhotau}
\eea
where $P_i$'s are the following projectors
\bea
({P_{TT}})^{\mu\nu\rho\tau}&=&R^{\mu\nu}R^{\rho\tau}~, \qquad \qquad \qquad
     ({P^a_{TT}})^{\mu\nu\rho\tau}=R^{\mu\rho}R^{\nu\tau}-R^{\mu\tau}R^{\nu\rho}~,
\label{PaTT} \nonumber \\
          ({P^{\tau}_{TT}})^{\mu\nu\rho\tau}&=&\frac{1}{2}\left[
R^{\mu\rho}R^{\nu\tau}+R^{\mu\tau}R^{\nu\rho}-R^{\mu\nu}R^{\rho\tau}\right]~, \qquad
({P_{ST}})^{\mu\nu\rho\tau}=k_1^{\mu}k_1^{\nu}R^{\rho\tau}~,  \nonumber \\
   ({P_{TS}})^{\mu\nu\rho\tau}&=&R^{\mu\nu}k_2^{\rho}k_2^{\tau}~, \qquad \qquad
\qquad
({P_{SS}})^{\mu\nu\rho\tau}=k_1^{\mu}k_1^{\nu}k_2^{\rho}k_2^{\tau}~,
\label{PST}
\\ ({P^{\tau}_{TS}})^{\mu\nu\rho\tau}&=&R^{\mu\rho}k_1^{\nu}k_2^{\tau}
+ R^{\mu\tau}k_1^{\nu}k_2^{\rho}+k_1^{\mu}k_2^{\rho}R^{\nu\tau}+
k_1^{\mu}k_2^{\tau}R^{\nu\rho}, \nonumber\\
({P^{\tau a}_{TS}})^{\mu\nu\rho\tau}&=&R^{\mu\rho}k_1^{\nu}k_2^{\tau}
- R^{\mu\tau}k_1^{\nu}k_2^{\rho}+k_1^{\mu}k_2^{\rho}R^{\nu\tau}
-k_1^{\mu}k_2^{\tau}R^{\nu\rho}~, \label{PtauaTS}\nonumber
\eea
with
\bea
R^{\mu\nu}&=&-g^{\mu\nu}+\frac{1}{X}\left[w
\left(q^{\mu}p^{\nu}+p^{\mu}q^{\nu}
\right) -q^2p^{\mu}p^{\nu}-p^2q^{\mu}q^{\nu}
\right]~,  \nonumber\\
k^{\mu}_1&=&\sqrt{\frac{-q^2}{X}}\left( p^{\mu}-\frac{w}{q^2}q^{\mu}  \right)
~, \qquad
k^{\mu}_2=\sqrt{\frac{-p^2}{X}}\left( q^{\mu}-\frac{w}{p^2}p^{\mu}  \right)
\eea
and \ $w=p\cdot q$ and\   $X=(p\cdot q)^2-p^2q^2$~.
Note that $R_{\mu\nu}$ is the metric tensor of the subspace which is
orthogonal
to $q$ and $p$, and thus $k_1^{\mu}R_{\mu\nu}= k_2^{\mu}R_{\mu\nu}=0$~.
Some useful properties of the projectors are given in Appendix A.
The virtual photon structure functions $W_i$ are functions of
three invariants, i.e., $w$, $q^2(=-Q^2)$ and $p^2(=-P^2)$, and have no
kinematical singularities.
The subscript \lq\lq T\rq\rq\  and \lq\lq S\rq\rq\  refer
to the transverse and longitudinal photon, respectively.
The structure functions with the superscript \lq\lq $\tau$\rq\rq\
correspond to transitions with spin-flip for each of the photons with
total helicity conservation, while those with the superscript  \lq\lq 
$a$\rq\rq\
correspond to the $\mu\nu$ antisymmetric part of $W^{\mu\nu\rho\tau}$
and are measured, for example,  through  the two-photon processes in polarized
$e^+e^-$ collision experiments.
These eight structure functions
are related to the $s$-channel helicity amplitudes as follows~\cite{BCG}:
\bea
W_{TT}&=&\frac{1}{2}\left[W(1,1\vert 1,1)+W(1,-1\vert 1,-1)   \right]~,
\qquad W_{ST}= W(0,1\vert 0,1)~,  \nonumber \\
W_{TS}&=&W(1,0\vert 1,0)~, \qquad W_{SS}=W(0,0\vert 0,0)~,    \nonumber \\
W^a_{TT}&=&\frac{1}{2}\left[W(1,1\vert 1,1)-W(1,-1\vert 1,-1)   \right]~,
\qquad W^{\tau}_{TT}= W(1,1\vert -1,-1)~,  \nonumber \\
W^{\tau}_{TS}&=& \frac{1}{2}\left[W(1,1\vert 0,0)-W(1,0\vert 0, -1)   \right]~,
\nonumber
\\ W^{\tau a}_{TS}&=&\frac{1}{2}\left[W(1,1\vert 0,0)+W(1,0\vert 0, -1)
\right]~. \label{HelicityVSBCG}
\eea
Since the helicity-nonflip amplitudes are non-negative, the first four
structure functions are  positive definite and the last four are not. Due to the
fact that the
absorptive part $W^{\mu\nu\rho\tau}(p,q)$  is symmetric under the simultaneous
interchange of $\{q,\mu,\nu\}\leftrightarrow \{p,\rho,\tau\}$, all the
virtual photon structure functions, except $W_{ST}$ and $W_{TS}$, are symmetric
under interchange of $p \leftrightarrow q$, while $W_{ST}(w,q^2,p^2)
=W_{TS}(w,p^2,q^2)$.
In terms of these structure functions, the positivity constraints
(\ref{CS1})-(\ref{CS3}) are rewritten as
\bea
\Bigl|W_{ TT}^\tau \Bigr|&\leq&
\left(W_{ TT}+W_{ TT}^a\right)~,  \label{BCG1}\\
\Bigl|W_{ TS}^\tau +W_{ TS}^{\tau a}  \Bigr|&\leq&
\sqrt{(W_{ TT}+W_{ TT}^a)W_{ SS}}~,\label{BCG2}\\
\Bigl| W_{ TS}^\tau -W_{ TS}^{\tau a} \Bigr|&\leq&
\sqrt{W_{ TS}W_{ ST}}\label{BCG3}~.
\eea
In fact,  the following bounds,
\be
         \Bigl|W_{ TT}^\tau \Bigr|\leq 2 W_{ TT}~, \qquad
2\Bigl( W_{ TS}^\tau  \Bigr)^2 \leq 2W_{SS} W_{ TT}+W_{ TS}W_{ ST}
\label{oldCostraint}
\ee
were derived, some time ago, from the positiveness of the $\gamma\gamma$
cross-section for arbitrary  photon polarization \cite{BGM}. Note that the
 constraints (\ref{BCG1})-(\ref{BCG3}) which we have obtained
are more stringent than the above ones (\ref{oldCostraint}).

\section{Parton Model Results}

For the real photon target, $P^2=0$, the number of independent
structure functions or helicity amplitudes reduces to four. They are
$W_{\rm TT}$, $W_{\rm ST}$, $W_{\rm TT}^\tau$, and $W_{\rm TT}^a$,
which are often referred to as
\be
W_{\rm TT}=W_1^\gamma~, \quad W_{\rm ST}=\frac{1}{2x}F_L^\gamma~, \quad
W_{\rm TT}^\tau=2W_3^\gamma~, \quad W_{\rm TT}^a=W_4^\gamma~,
\ee
and we have only one
positivity constraint (\ref{BCG1}). In our previous paper~\cite{SSU} we have
examined this constraint in the simple PM.
Up to now most of our attention has been focused on the study of these four
functions. In the case of virtual photon target, $P^2\neq 0$, there appear
four additional   structure functions and we have derived three
positivity constraints. But since
we do not have much knowledge on the new photon structure functions, it is
worthwhile, first, to investigate these functions in the simple PM and then to 
examine
whether the three positivity constraints (\ref{BCG1})-(\ref{BCG3}) actually
hold.

We evaluate the box (massive quark-loop with a quark mass $m$) diagrams shown
in Fig. 2.
By applying the  projectors, which were given in (\ref{PST}), to the box
diagram
contributions,  we obtain the PM predictions for the eight virtual photon
structure
functions,
$W_{TT}\vert_{PM}$, $W_{TT}^a\vert_{PM}$, $W_{TT}^{\tau}\vert_{PM}$,
$W_{ST}\vert_{PM}$,
$W_{TS}\vert_{PM}$, $W_{SS}\vert_{PM}$, $W_{TS}^{\tau}\vert_{PM}$, and
$W_{TS}^{\tau a}\vert_{PM}$. Their explicit expressions for the case
$m\neq 0$ and $P^2\neq 0$  are given in Appendix B.2.
The results are consistent with the cross sections for the
$\gamma\gamma \rightarrow e^+e^-(\mu^+\mu^-)$ process obtained by Budnev
et al.~\cite{BGMS} except for $W_{TS}^{\tau a}\vert_{PM}$
\footnote{In the expression of $\tau^a_{TS}$, the last one in Eq.(E.1) of
Ref.\cite{BGMS} which corresponds to our $W_{TS}^{\tau a}\vert_{PM}$, the
factor
$\left[L+\frac{(q_1q_2)\Delta t}{T}\right]$  should read as
$\left[L-\frac{(q_1q_2)\Delta t}{T}\right]$.}. Also the expressions of
$W_{TT}\vert_{PM}$, $W_{ST}\vert_{PM}$, and $W_{SS}\vert_{PM}$
are, respectively, in accord with those of $F_{TT}$, $F_{LT}$, and $F_{SS}$
given in Ref.{\cite{GRSch}}.

We plot, in Fig.3 (a)-(b), these PM results for the eight  photon structure
functions
as functions of $x=Q^2/(2p\cdot q)$. The vertical axes are in units
\footnote{Our definition of $W_{\mu\nu\rho\tau}$ and therefore of the
photon structure functions, is such that they are proportional to
$\alpha=e^2/4\pi$, and not to $\alpha^2$, in conformity with the nucleon case. }
of $(\alpha/2\pi)\delta_\gamma$, where $\delta_\gamma=3\sum_i^{N_f}e_i^4$~,
with
$N_f$ the number of active flavors. We have taken $P^2/Q^2=1/30$ and
$m^2/Q^2=1/100$.
The allowed $x$ region is $0\le x \le x_{\rm max}$ with
\be
    x_{\rm max} =1/\left( 1+\frac{P^2}{Q^2} +\frac{4m^2}{Q^2}  \right)~.
\ee

  From Fig.3 (a)-(b), we see that the photon structure functions can be
classified into three groups according to
their magnitude: $W_{TT}\vert_{PM}$ and $W_{TT}^a\vert_{PM}$ are
the first group,
$W_{ST}\vert_{PM}$, $W_{TT}^{\tau}\vert_{PM}$, $W_{TS}\vert_{PM}$, and
$W_{TS}^{\tau}\vert_{PM}$ are the second one and $W_{SS}\vert_{PM}$ and
$W_{TS}^{\tau a}\vert_{PM}$ are the third one. By comparison with $W_{TT}\vert_{PM}$ and
$W_{TT}^a\vert_{PM}$ in the first group, $W_{SS}\vert_{PM}$ and
$W_{TS}^{\tau a}\vert_{PM}$ are extremely small in magnitude.
Also we see that the helicity-flip structure functions
$W_{TT}^{\tau}\vert_{PM}$ and
$W_{TS}^{\tau}\vert_{PM}$ are smaller in magnitude than the helicity-nonflip ones
$W_{TT}\vert_{PM}$ and $W_{TS}\vert_{PM}$, respectively. We expect that these
characteristics of the PM results will persist in the actual photon
structure functions which would be obtained from future experiments.

The graphs in Fig.3 (a)-(b) show that all photon stucture functions tend to
vanish as
$x\rightarrow x_{\rm max}$ and this is the consequence of the kinematical
constraint.
For $x\rightarrow 0$, $W_{TT}\vert_{PM}$ and $W_{TT}^a\vert_{PM}$  both
diverge. The former
diverges positively, and the later negatively. However the sum remains finite since
$(W_{TT}\vert_{PM}+W_{TT}^a\vert_{PM})/(\frac{\alpha}{2\pi}\delta_\gamma
)\rightarrow 2$ as
$x\rightarrow 0$ (see Fig.5 (a) below). The other structure functions
vanish at $x=0$.

It is interesting to note the clear difference, in the PM predictions
for the photon structure functions, between the massive and the 
massless quark cases.
We plot in Fig.4 (a)-(b) the results for the massless quark case, $m=0$,
with
$P^2/Q^2=1/30$. Now  three structure functions, $W_{TT}\vert_{PM}$,
$W_{TT}^a\vert_{PM}$, and
$W_{TT}^{\tau}\vert_{PM}$, do not vanish as $x\rightarrow x_{\rm max}$ and
remain finite.
For $x\rightarrow 0$, $W_{TT}\vert_{PM}$ and $W_{TT}^a\vert_{PM}$  both
diverge again, but the sum tends to zero (see Fig.6 (a)  below).

From the symmetry argument on the absorptive part $W^{\mu\nu\rho\tau}(p,q)$,
we know that $W_{ST}$
and $W_{TS}$  switch into one another under interchange of $p
\leftrightarrow q$, namely,
$W_{ST}(w,q^2,p^2) =W_{TS}(w,p^2,q^2)$.  But this does not mean
$W_{ST}=W_{TS}$. Indeed according to the PM
results in the massive quark case shown in Figs.3 (a)-(b),
 $W_{ST}\vert_{PM}$ and $W_{TS}\vert_{PM}$ are different in
magnitude and also
have different $x$-dependences.
However, we have found that in the limit $m=0$, $W_{ST}\vert_{PM}$
coincides with  $W_{TS}\vert_{PM}$
irrespectively to the values of $P^2$ and $Q^2$, which we believe is not a
trivial result.

We plot in Fig.5 (a) the PM predictions versus $x$ 
of $(W_{\rm TT}+W_{\rm TT}^a)$ and
$\Bigl|W_{\rm TT}^\tau\Bigr|$, in Fig.5 (b) those of
 $\sqrt{(W_{\rm TT}+W_{\rm TT}^a)W_{\rm SS}}$ and $\Bigl|W_{\rm TS}^\tau
+W_{\rm TS}^{\tau
a}  \Bigr|$, and in Fig.5 (c) those of $\sqrt{W_{\rm TS}W_{\rm ST}}$ and $\Bigl| W_{\rm
TS}^\tau -W_{\rm TS}^{\tau a}\Bigr|$,  for the case  $P^2/Q^2=1/30$ and
$m^2/Q^2=1/100$.  For massless quark $m=0$, with
$P^2/Q^2=1/30$, similar plots are shown in Figs.6 (a)-(c).
In both cases we can see that the three positivity constraints
(\ref{BCG1})-(\ref{BCG3}) are
indeed satisfied for all the allowed $x$ region. However, as we have already
mentioned above,
the behaviors of the sum $(W_{\rm TT}+W_{\rm TT}^a)_{PM}$ show a clear
difference between the
massive and the massless quark cases (see Fig.5 (a) and Fig.6 (a)). For massive
quark,
the sum reaches $2\times (\frac{\alpha}{2\pi}\delta_\gamma)$  as
$x\rightarrow 0$ and the positivity
constraint (\ref{BCG1}) is satisfied  for all the allowed $x$ region with a
wide margin. On the other
hand, for massless quark, it vanishes as $x\rightarrow 0$ and the
difference between
$(W_{\rm TT}+W_{\rm TT}^a)_{PM}$ and $\Bigl|W_{\rm TT}^\tau\Bigr|_{PM}$
reduces to zero.
The fact that the sum $(W_{\rm TT}+W_{\rm TT}^a)_{PM}$ vanishes at $x=0$
in the case of massless quark is explained as follows. The sum is related
to a $s$-channel helicity
amplitude of
$\gamma$-$\gamma$ scattering, $W_{\rm TT}+W_{\rm TT}^a=W(1,1\vert 1,1)$.
Now the limiting procedure $x=\frac{Q^2}{2p\cdot q} \rightarrow 0$ with the
ratio $\frac{P^2}{Q^2}$ fixed,
is equivalent of taking  $P^2\rightarrow 0$ and $Q^2\rightarrow 0$ and
keeping $2p\cdot q$ finite.
So the situation at $x=0$ is the same as if we were dealing with  the two {\it 
real}
photon scattering process, $\gamma+\gamma \rightarrow q+{\overline q}$. 
Since 
chirality coincides with helicity for massless quark and the electromagnetic
interaction preserves the
quark chirality, it is known that the amplitude for the two real photons
with the same
helicity annihilating into a massless quark pair, vanishes identically
~\cite{Parke}.


\section {Perturbative QCD}

Now we switch on the QCD coupling. The photon structure functions have been
studied by pQCD for many years~\cite{MK}. Especially,
in the kinematic region,
\be
\Lambda^2\ll P^2 \ll Q^2  \label{KinRegion}
\ee
where the mass squared of the target photon ($P^2$) is much larger than the
QCD scale parameter ($\Lambda^2$), some of the photon
structure functions are predictable in pQCD entirely up to the
NLO. This is due to the fact that, in this kinematical
region, the
hadronic components of the photon  (in other words, the photon matrix elements of
hadronic operators) can  be calculated perturbatively.
Indeed, the  virtual photon structure functions, such as unpolarized
$F_2^\gamma (x,Q^2,P^2)$ and $F_L^\gamma (x,Q^2,P^2)$~\cite{UW2} and  polarized
$g_1^\gamma (x,Q^2,P^2)$~\cite{SU}, were studied up to the NLO
in the above kinematic region.
Here the virtual photon structure functions $F_2^\gamma$, $F_L^\gamma$, and 
$g_1^\gamma$
are related to the ones  introduced by BCG  in (\ref{Wmunurhotau}) as 
follows
\footnote{We follow Nisius \cite{Nisius} for the definition of $F_1^\gamma$,
$F_2^\gamma$, and $F_L^\gamma$ apart from $F_1^\gamma$ being different from
the one of Nisius by a factor of 2. For other definitions of $F_1^\gamma$,
$F_2^\gamma$, and $F_L^\gamma$, see Refs.\cite{GRS,BW}.}:
\bea
F_1^\gamma(x,Q^2,P^2)&=&2\left[ W_{TT}-\frac{1}{2}W_{TS}\right], \nonumber \\
F_2^\gamma(x,Q^2,P^2)&=&\frac{2x}{{\widetilde\beta}^2} \left[
   W_{TT}+W_{ST}-\frac{1}{2}W_{SS} -\frac{1}{2}W_{TS}  \right], \nonumber \\
F_L^\gamma(x,Q^2,P^2)&=&F_2^\gamma-xF_1^\gamma,   \\
g_1^\gamma(x,Q^2,P^2)&=&\frac{2}{{\widetilde\beta}^2}\left[W^a_{TT}-
\frac{(P^2Q^2)^{1/2}}{w}W^{\tau a}_{TS}
     \right],  \nonumber
\eea
with ${\widetilde\beta}=\left(1-\frac{P^2Q^2}{w^2}\right)^{1/2}$.

Since the tensor $W^{\mu\nu\rho\tau}(p,q)$ in (\ref{Wmunurhotau}) is regular
as $p^2 \rightarrow 0$, while the projectors $P_{TS}$ and $P^{\tau a}_{TS}$ 
are singular
as $p^2 \rightarrow 0$ and behave as $1/p^2$ and $1/\sqrt{-p^2}$, respectively,
we expect $W_{TS}\propto \frac{P^2}{Q^2}$ and $W^{\tau a}_{TS}\propto
\sqrt{\frac{P^2}{Q^2}}$. Then, in the  kinematic region (\ref{KinRegion}),
$\widetilde\beta  \approx 1$, and we can neglect the contributions of
$W_{TS}$  and $W^{\tau a}_{TS}$  as compared with $W_{TT}$ and $W^a_{TT}$, 
respectively.
As a result we have
\bea
W_{TT}(x,Q^2,P^2)&\approx&\frac{1}{2} F_1^\gamma(x,Q^2,P^2)=
\frac{1}{2x}\left\{F_2^\gamma(x,Q^2,P^2)-F_L^\gamma(x,Q^2,P^2)\right\},
\nonumber \\
W^a_{TT}(x,Q^2,P^2)&\approx& \frac{1}{2}g_1^\gamma(x,Q^2,P^2)~.
\eea
The positivity constraint (\ref{BCG1}) is now rewritten as
\be
\Bigl|W_{ TT}^\tau(x,Q^2,P^2) \Bigr|
\mathop{<}_\sim
\frac{1}{2}\left[F_1^\gamma(x,Q^2,P^2)+g_1^\gamma(x,Q^2,P^2)\right]~, 
\label{newBCG1}
\ee
and it is interesting to see if this inequality is satisfied by the pQCD 
results.
For $F_2^\gamma (x,Q^2,P^2)$ and $F_L^\gamma (x,Q^2,P^2)$, we can take the
results from Ref.\cite{UW2} and for $g_1^\gamma (x,Q^2,P^2)$ we use
Ref.\cite{SU}.  Actually the pQCD results for $F_1^\gamma$ and $g_1^\gamma$ are
given in the form of Mellin moments, and we need to perform the inverse
Mellin transformation in order to express them as functions of $x$.  The 
formula for the
$n$-th  moment of $F_1^\gamma$ up to the NLO is summarized in Appendix C.
After the inverse Mellin transformation,  $\frac{1}{2} 
(F_1^\gamma+g_1^\gamma)$ is
expressed in the form as
\be
\frac{1}{2} (F_1^\gamma+g_1^\gamma)=\frac{\alpha}{2\pi}\delta_{\gamma}
\left[ a(x) {\rm ln}\frac{Q^2}{\Lambda^2}+b(x)+{\cal O}(\alpha_s(Q^2))  \right],
\ee
where $\Lambda^2$ is the QCD scale and $\alpha_s(Q^2)$ is the QCD running
coupling constant.

The virtual photon structure function $W_{TT}^\tau(x,Q^2,P^2)(=2W_3^\gamma)$
is expected to be given by the same expression as the PM result
up to  ${\cal O} (1/\ln (Q^2/\Lambda^2))$, since there exist no twist-2 
quark operators
contributing to $W_{TT}^\tau$. So we take in the leading order 
(LO)~\cite{KS,MANO},
\be
W_{TT}^\tau(x,Q^2,P^2)=\frac{\alpha}{2\pi}\delta_{\gamma}
\left[(-2x^2)  +{\cal O}(\alpha_s(Q^2)) \right],
\ee
where the first term is derived from $W^{\tau}_{TT}\vert_{PM}$ given in
(\ref{WTTtau}), ignoring the power corrections of $m^2/Q^2$ and $P^2/Q^2$.

Now we plot, in Fig. 7, the NLO pQCD result of $\frac{1}{2} 
(F_1^\gamma+g_1^\gamma)$
and the LO result of $\Big|W_{TT}^\tau\Big|$ as functions of $x$ for the case
$P^2/Q^2=1/30$  with the number of active flavors, $N_f=3$. We find that 
the inequality
(\ref{newBCG1}) is satisfied for almost all the allowed $x$ region except near
$x_{\rm max}=\frac{Q^2}{Q^2+P^2}(\approx 0.968\  {\rm for}\ 
P^2/Q^2=1/30)$. The violation
of the inequality  near $x_{\rm max}$ is explained as follows.
We observe that the graph of $\frac{1}{2}
(F_1^\gamma+g_1^\gamma)$ falls  rapidly as $x\rightarrow x_{\rm max}$.
In the language of the QCD improved parton model,
this is due to total momentum conservation of all partons in the photon.
In fact, the moments of both $F_1^\gamma$ and $g_1^\gamma$ in the LO
behave  as $1/(n{\rm ln}~n)$ for large $n$ and thus in $x$-space they
vanish like $-1/{\rm ln}(1-x) $ as $x\rightarrow 1$. The NLO QCD corrections
further suppress
$F_1^\gamma$ and $g_1^\gamma$ at large $x$.
On the other hand, the LO QCD prediction of $W_{TT}^\tau$ is the same as 
the massless
quark  PM result, with the power corrections of $P^2/Q^2$ being neglected.
Thus $\Big|W_{TT}^\tau\Big|$ in Fig.7 increases monotonically as a function 
of $x^2$
and the violation of the inequality (\ref{newBCG1}) occurred near $x_{\rm max}$.

However, the {\it physical} $W_{TT}^\tau$ should vanish as $x\to x_{\rm 
max}$ due to
kinematical constraints. The momentum conservation of  partons is not 
applicable
here since quark partons in the photon do not contribute to $W_{TT}^\tau$
in the LO, in other words, there exist no twist-2 quark operators
relevant to $W_{TT}^\tau$~\cite{KS}.  This urges the necessity of 
introducing the
quark mass effects to the calculation of the photon coefficient function.
(Remember that $W_{TT}^\tau|_{\rm PM}$ in the massive PM vanishes as 
$x\rightarrow
x_{\rm max}$.)

Except for large and small $x$, we find that the pQCD prediction for
$\frac{1}{2}(F_1^\gamma+g_1^\gamma)$ appears to be similar to the massless 
quark PM
result for $(W_{\rm TT}+W_{\rm TT}^a)$. In fact,
for moderate $x$, $0.2\leq x \leq 0.7$,
the graph
$\frac{1}{2}(F_1^\gamma+g_1^\gamma)$ resembles closely
the massless quark result
$(W_{\rm TT}+W_{\rm TT}^a)_{\rm PM}$ in Fig.6 (a) in shape and magnitude.
As $x\to 0$,    we find that the sum $\frac{1}{2}(F_1^\gamma+g_1^\gamma)$ 
starts to increase.

\section{Conclusion}

To summarize we have investigated the three positivity constraints
for the virtual photon structure functions which could be studied
in future $ep$ and $e^+e^-$ colliders. 
In particular the virtual photon structure can be measured from the 
double-tagged $e^+e^-$ reactions and also from the dijet events in 
deep inelastic $ep$ collisions.  

By evaluating the quark box-diagrams, we obtained the eight virtual
photon structure functions in the PM both for a massive and a massless quark. 
 It has turned out that there exists a clear difference
both in $x$-dependence and in magnitude between the massive and massless
quark PM predictions. We have found that the three constraints indeed 
hold for the PM computation of both massive and massless quark cases. 
In the kinematic region,  
$\Lambda^2 \ll P^2 \ll Q^2$, the NLO QCD results for $F_1^\gamma$ and 
$g_1^\gamma$ and the LO result for $W^\tau_{\rm TT}$ satisfy the
constraint among these three structure functions 
for most of the allowed $x$ region except for the region very
near $x_{\rm max}$.

We expect that these bounds will provide useful constraints 
for studying the yet unknown polarized and unpolarized virtual 
photon structure functions.


\vspace{1cm}

\vspace{0.5cm}
\leftline{\large\bf Acknowledgement}
\vspace{0.5cm}

We thank Werner Vogelsang and Hideshi Baba for useful discussions.
This work is partially supported by the Grant-in-Aid
from the Japan Ministry of Education and Science,
No.(C)(2)-12640266.


\newpage
\appendix
\section{Projectors for the virtual photon structure functions}

The projectors $P_i$'s are defined in (\ref{PST}).
$ ({P^a_{TT}})^{\mu\nu\rho\tau}$ and $({P^{\tau a}_{TS}})^{\mu\nu\rho\tau}$
   are
antisymmetric under the interchange of $\mu
\leftrightarrow   \nu$ and
$\rho \leftrightarrow   \tau$, while other projectors are symmetric.
Since $R_{\mu\nu}$, $k_1$, and $k_2$ have the following properties,
\bea
R_{\mu\nu}q^{\mu}&=&0~, \quad R_{\mu\nu}p^{\mu}=0~, \quad
R_{\mu\rho}{R^{\rho}}_{\nu}=-R_{\mu\nu}~, \quad R_{\mu\nu}R^{\mu\nu}=2~,
\nonumber\\
k_1^{\mu}R_{\mu\nu}&=&0~, \qquad  k_2^{\mu}R_{\mu\nu}=0~, \qquad
k_1^2=k_2^2=1~.
\eea
we find
\bea
({P_i})^{\mu\nu\rho\tau}({P_j})_{\mu\nu\rho\tau}&=&0 \qquad {\rm for}\qquad
i\neq
j\nonumber\\ ({P_{TT}})^{\mu\nu\rho\tau}({P_{TT}})_{\mu\nu\rho\tau}&=&4~,
\qquad
({P^a_{TT}})^{\mu\nu\rho\tau}({P^a_{TT}})_{\mu\nu\rho\tau}=4~, \nonumber\\
({P^{\tau}_{TT}})^{\mu\nu\rho\tau}({P^{\tau}_{TT}})_{\mu\nu\rho\tau}&=&2~,
\qquad
({P_{ST}})^{\mu\nu\rho\tau}({P_{ST}})_{\mu\nu\rho\tau}=2~, \\
({P_{TS}})^{\mu\nu\rho\tau}({P_{TS}})_{\mu\nu\rho\tau}&=&2~, \qquad
({P_{SS}})^{\mu\nu\rho\tau}({P_{SS}})_{\mu\nu\rho\tau}=1~, \nonumber\\
({P^{\tau}_{TS}})^{\mu\nu\rho\tau}({P^{\tau}_{TS}})_{\mu\nu\rho\tau}&=&8~,
\qquad
({P^{\tau a}_{TS}})^{\mu\nu\rho\tau}({P^{\tau a}_{TS}})_{\mu\nu\rho\tau}=8~.
\nonumber
\eea

\section{Virtual photon structure functions in parton model}

\subsection{Parameters}

\bea
x&=&\frac{Q^2}{2p\cdot q}~, \qquad \delta_{\gamma}=3\sum_{i=1}^{N_f}e_i^4~,
\nonumber \\
{\widetilde\beta}&=&\sqrt{1-\frac{P^2Q^2}{(p\cdot q)^2}}~, \qquad
\beta=\sqrt{1-\frac{4m^2}{(p+q)^2}}~, \nonumber \\
L&=&{\rm ln}~\frac{1+\beta {\widetilde\beta}}{1-\beta {\widetilde\beta}}~,
\eea

where $N_f$ is the number of the active flavors and $m$ is the quark mass.

\subsection{Structure functions in PM}

\bea
W_{TT}\vert_{PM}&=&\frac{\alpha}{2\pi}\delta_{\gamma}\Biggl[
L\Biggl\{-\frac{8x^2}{{\widetilde\beta}}\frac{m^4}{Q^4}
+\frac{ {\widetilde\beta}^2-(1-2x)^2
}{{\widetilde\beta}^3}~\frac{m^2}{Q^2} \nonumber \\
&&\qquad  +\frac{1}{8{\widetilde\beta}^5}
\left[ (1-{\widetilde\beta}^2)({\widetilde\beta}^4+3)
-8x\left\{{\widetilde\beta}^4-2{\widetilde\beta}^2+3\right\}
\right]\frac{P^2}{Q^2}
\nonumber \\
&&\qquad +\frac{1}{4{\widetilde\beta}^5}\left[ -{\widetilde\beta}^6
+(2x^2-4x+7){\widetilde\beta}^4+(8x-11){\widetilde\beta}^2+3(2x^2-4x+3)
\right]
\Biggr\} \nonumber \\
&+&\frac{\beta}{1-\beta^2 {\widetilde\beta}^2}
\Biggl\{-16x^2\frac{m^4}{Q^4}
+\frac{2\left[(1-2x)^2+(4x-1){\widetilde\beta}^2
\right]}{{\widetilde\beta}^2}\frac{m^2}{Q^2}  \\
&&\qquad +\frac{1}{4{\widetilde\beta}^4}
\left[(1-{\widetilde\beta}^2)
({\widetilde\beta}^4+2\beta^2{\widetilde\beta}^2-3)
+8x\left\{\beta^2{\widetilde\beta}^4-2(\beta^2+1){\widetilde\beta}^2+3\right\}
\right]\frac{P^2}{Q^2}
\nonumber \\ &&\qquad  +\frac{1}{2{\widetilde\beta}^4}\biggl[
(2\beta^2+1){\widetilde\beta}^6+\left\{2x^2+4\beta^2x-6\beta^2-5
\right\}{\widetilde\beta}^4 \nonumber \\
&&\qquad  \quad+\left\{4\beta^2 x^2-8(\beta^2+1)x+6\beta^2+11
\right\}{\widetilde\beta}^2-3(2x^2-4x+3)
\biggr]
\Biggr\}
\Biggr]~, \label{WTT} \nonumber \\
&&   \nonumber \\
W^{a}_{TT}\vert_{PM}&=&\frac{\alpha}{2\pi}\delta_{\gamma}\Biggl[
L~\frac{1}{{\widetilde\beta}^3}\biggl\{2x\frac{P^2}{Q^2}+({\widetilde\beta}^2
+2x-2)
\biggr\} \nonumber \\
&&+\frac{\beta}{1-\beta^2 {\widetilde\beta}^2}
\biggl\{8x~\frac{m^2}{Q^2}
+\frac{2x\left\{(\beta^2+1){\widetilde\beta}^2-2\right\}}{{\widetilde\beta}^2}
\frac{P^2}{Q^2} \nonumber \\
&&\qquad \qquad \quad+\frac{({\widetilde\beta}^2+2x-2)
\left\{(\beta^2+1){\widetilde\beta}^2-2\right\}}{{\widetilde\beta}^2}
\biggr\}
\Biggr], \label{WTTa} \\
&&   \nonumber \\
W^{\tau}_{TT}\vert_{PM}&=&\frac{\alpha}{2\pi}\delta_{\gamma}\Biggl[
2L\Biggl\{-\frac{4x^2}{{\widetilde\beta}}\frac{m^4}{Q^4}
+\frac{ {\widetilde\beta}^4-4x(x+1){\widetilde\beta}^2-(1-2x)^2
}{2{\widetilde\beta}^3}~\frac{m^2}{Q^2} \nonumber \\
&&\qquad \quad +\frac{1-{\widetilde\beta}^2}{16{\widetilde\beta}^5}
\left[ (1-{\widetilde\beta}^2)(3+{\widetilde\beta}^2)
+8x({\widetilde\beta}^2-3)  \right]\frac{P^2}{Q^2} \nonumber \\
&&\qquad  \quad+\frac{1-{\widetilde\beta}^2}{8{\widetilde\beta}^5}\left[
3{\widetilde\beta}^4+2(x^2+2x-4){\widetilde\beta}^2+3(2x^2-4x+3)
\right]
\Biggr\} \nonumber \\
&&+\frac{2\beta}{1-\beta^2 {\widetilde\beta}^2}
\Biggl\{-8x^2\frac{m^4}{Q^4}
+\frac{(1-{\widetilde\beta}^2)\left[(1-2x)^2-{\widetilde\beta}^2
\right]}{{\widetilde\beta}^2}\frac{m^2}{Q^2}  \\
&&\qquad -\frac{1-{\widetilde\beta}^2}{8{\widetilde\beta}^4}
\left[{\widetilde\beta}^4-2(\beta^2+1){\widetilde\beta}^2+3
+8x\left\{(2\beta^2+1){\widetilde\beta}^2-3\right\}  \right]\frac{P^2}{Q^2}
\nonumber \\
&&\qquad  +\frac{1}{4{\widetilde\beta}^4}\biggl[
(4\beta^2+1){\widetilde\beta}^6-\left\{2x^2-4(2\beta^2+1)x+10\beta^2+9
\right\}{\widetilde\beta}^4 \nonumber \\
&&\qquad  \quad+\left\{4(\beta^2+1)x^2-8(\beta^2+2)x+6\beta^2+17
\right\}{\widetilde\beta}^2-3(2x^2-4x+3)
\biggr]
\Biggr\}
\Biggr]~,   \label{WTTtau} \nonumber \\
&&   \nonumber \\
W_{ST}\vert_{PM}&=&\frac{\alpha}{2\pi}\delta_{\gamma}\Biggl[
L\Biggl\{
-\frac{ {\widetilde\beta}^4+(4x^2+4x-2){\widetilde\beta}^2+(1-2x)^2
}{{\widetilde\beta}^3}~\frac{m^2}{Q^2} \nonumber \\
&&\qquad \quad +\frac{1-{\widetilde\beta}^2}{4{\widetilde\beta}^5}
\left[ (1-{\widetilde\beta}^2)(3+{\widetilde\beta}^2)
-24x  \right]\frac{P^2}{Q^2} \nonumber \\
&&\qquad  \quad-\frac{1-{\widetilde\beta}^2}{2{\widetilde\beta}^5}\left[
{\widetilde\beta}^4-2(x^2-2){\widetilde\beta}^2-3(2x^2-4x+3)
\right]
\Biggr\} \nonumber \\
&&+\frac{\beta}{1-\beta^2 {\widetilde\beta}^2}
\Biggl\{
\frac{2(1-{\widetilde\beta}^2)\left[(1-2x)^2-{\widetilde\beta}^2
\right]}{{\widetilde\beta}^2}\frac{m^2}{Q^2}  \\
&&\qquad +\frac{1-{\widetilde\beta}^2}{2{\widetilde\beta}^4}
\left[(1-{\widetilde\beta}^2)\left\{
(2\beta^2-1){\widetilde\beta}^2-3 \right\}
-8x~(2\beta^2{\widetilde\beta}^2-3)  \right]\frac{P^2}{Q^2} \nonumber \\
&&\qquad  +\frac{1}{{\widetilde\beta}^4}\biggl[
\beta^2 {\widetilde\beta}^6-\left\{2x^2-4(\beta^2+1)x+7\beta^2+4
\right\}{\widetilde\beta}^4 \nonumber \\
&&\qquad  \quad+\left\{4(\beta^2+1)x^2-4(2\beta^2+3)x+6\beta^2+13
\right\}{\widetilde\beta}^2-3(2x^2-4x+3)
\biggr]
\Biggr\}
\Biggr]~,    \label{WST} \nonumber \\
&&   \nonumber \\
W_{TS}\vert_{PM}&=&\frac{\alpha}{2\pi}\delta_{\gamma}\Biggl[
L~(1-{\widetilde\beta}^2)\Biggl\{
-\frac{ 4x^2-4x+{\widetilde\beta}^2+1
}{{\widetilde\beta}^3}~\frac{m^2}{Q^2} \nonumber \\
&&\qquad \qquad \qquad +\frac{1}{4{\widetilde\beta}^5}
\left[ (1-{\widetilde\beta}^2)(3+{\widetilde\beta}^2)
-24x  \right]\frac{P^2}{Q^2} \nonumber \\
&&\qquad  \qquad \qquad -\frac{1}{2{\widetilde\beta}^5}\left[
{\widetilde\beta}^4-2(x^2-2){\widetilde\beta}^2-3(2x^2-4x+3)
\right]
\Biggr\} \nonumber \\
&&+\frac{\beta}{1-\beta^2 {\widetilde\beta}^2}
\Biggl\{
\frac{2(1-{\widetilde\beta}^2)\left[(1-2x)^2-{\widetilde\beta}^2
\right]}{{\widetilde\beta}^2}\frac{m^2}{Q^2}  \\
&&\qquad +\frac{1}{2{\widetilde\beta}^4}
\biggl[-(1-{\widetilde\beta}^2)\left\{{\widetilde\beta}^4
-2(\beta^2+1){\widetilde\beta}^2+3 \right\} \nonumber \\
&&\qquad \qquad \qquad \qquad \quad+8x \left\{
(\beta^2+1){\widetilde\beta}^4-(2\beta^2+3){\widetilde\beta}^2+3 \right\}
\biggr]\frac{P^2}{Q^2} \nonumber \\
   &&\qquad  -\frac{1-{\widetilde\beta}^2}{{\widetilde\beta}^4}\biggl[
\beta^2 {\widetilde\beta}^4-2\left\{(2\beta^2-1) x^2-4\beta^2x+3\beta^2+2
\right\}{\widetilde\beta}^2 \nonumber \\
&&\qquad \qquad \qquad \qquad \qquad \qquad \qquad \qquad \qquad
+3(2x^2-4x+3)
\biggr]
\Biggr\}
\Biggr]~,      \label{WTS} \nonumber \\
&&   \nonumber \\
W_{SS}\vert_{PM}&=&\frac{\alpha}{2\pi}\delta_{\gamma}\Biggl[
L~\frac{(1-{\widetilde\beta}^2)(3-{\widetilde\beta}^2)}{2{\widetilde\beta}^5}
-\frac{\beta~(1-{\widetilde\beta}^2)
\left\{3-(2\beta^2+1) {\widetilde\beta}^2  \right\}}
{(1-\beta^2 {\widetilde\beta}^2){\widetilde\beta}^4}\Biggr]
\nonumber \\
&&\qquad \qquad \times \biggl\{
\left[ (1-{\widetilde\beta}^2)-8x \right]\frac{P^2}{Q^2} +
2(2x^2-4x+3-{\widetilde\beta}^2)
\biggr\}~,  \label{WSS}  \\
&&   \nonumber \\
W^{\tau}_{TS}\vert_{PM}&=&-\frac{\alpha}{2\pi}\delta_{\gamma}\Biggl[
L~\frac{\sqrt{1-{\widetilde\beta}^2}}{{\widetilde\beta}^5}
\Biggl\{{\widetilde\beta}^2\left[{\widetilde\beta}^2-(1-2x)^2
\right]\frac{m^2}{Q^2}
\nonumber \\
&&\qquad \qquad \quad+\left[
\frac{1}{4}(1-{\widetilde\beta}^2)(3-{\widetilde\beta}^2)
+2x(2{\widetilde\beta}^2-3)\right]\frac{P^2}{Q^2}
\nonumber \\
&&\qquad \qquad
\quad+\frac{1}{2}\left[{\widetilde\beta}^4-2(x^2-4x+5){\widetilde\beta}^2
+3(2x^2-4x+3)\right]
\Biggr\} \nonumber \\
&+&\frac{\beta\sqrt{1-{\widetilde\beta}^2}}{{\widetilde\beta}^4(1-\beta^2
{\widetilde\beta}^2)}
\Biggl\{2{\widetilde\beta}^2(2x-1)\left(2x-1+{\widetilde\beta}^2\right)~\frac{m^2}{Q^2}
   \\
&&\quad +
\left[\frac{1}{2}(1-{\widetilde\beta}^2)\left\{(2\beta^2+1){\widetilde\beta}^2-3
\right\}+x\left\{(3\beta^2+1){\widetilde\beta}^4-8(\beta^2+1){\widetilde\beta}^2+12
\right\} \right]\frac{P^2}{Q^2}
   \nonumber \\
&&\quad+(3\beta^2x-5\beta^2+x-2){\widetilde\beta}^4
+2\left\{(2\beta^2+1)x^2-4(\beta^2+1)x+3\beta^2+5  \right\}{\widetilde\beta}^2
   \nonumber \\
&&\qquad \qquad \qquad \qquad \qquad \qquad \qquad -3(2x^2-4x+3)
\Biggr\}
\Biggr]~,  \label{WTStau} \nonumber \\
&&   \nonumber \\
W^{\tau a}_{TS}\vert_{PM}&=&\frac{\alpha}{2\pi}\delta_{\gamma}
~\frac{(1-{\widetilde\beta}^2)^{3/2}}{{\widetilde\beta}^3}
\Biggl[L-\frac{2\beta{\widetilde\beta} }{1-\beta^2
{\widetilde\beta}^2}\Biggr]\biggl\{1-x-x\frac{P^2}{Q^2} \biggr\}~.
\label{WTStaua}
\eea

\newpage

\section{The $n$-th moment of $F_1^\gamma$ in pQCD}

The pQCD prediction for the $n$-th moment of $F_1^\gamma$ up to NLO is
summarized as follows:
\bea
&& \int_0^1dx x^{n-1}F_1^\gamma(x,Q^2,P^2)=\int_0^1dx
x^{n-2}\left(F_2^\gamma-F_L^\gamma\right)\nonumber\\
&&=\frac{\alpha}{4\pi}
\frac{1}{2\beta_0}\left[
\sum_{i=+,-,NS}\tilde{P}_i^n\frac{1}{1+\lambda_i^n/2\beta_0}
\frac{4\pi}{\alpha_s(Q^2)}\left\{
1-\left(\frac{\alpha_s(Q^2)}{\alpha_s(P^2)}\right)^
{\lambda_i^n/2\beta_0+1}\right\}\right.\nonumber\\
&&\left.+\sum_{i=+,-,NS} A_i^n\left\{1-\left(\frac{\alpha_s(Q^2)}
{\alpha_s(P^2)}\right)^{\lambda_i^n/2\beta_0}\right\}\right.\nonumber\\
&&\left.+\sum_{i=+,-,NS} \left(B_i^n-\frac{\tilde{P}_i^{(L),n}}
{1+\lambda_i^n/2\beta_0}\right)\left\{1-\left(\frac{\alpha_s(Q^2)}
{\alpha_s(P^2)}\right)^{\lambda_i^n/2\beta_0+1}\right\}\right.\nonumber\\
&&\left.+C_\gamma^n-2\beta_0\delta_\gamma B_{\gamma,L}^n\right]~.
\eea
where $\beta_0(=11-\frac{2N_f}{3})$ is the one-loop QCD $\beta$ function and
$\alpha_s(Q^2)$ is the QCD running coupling constant.
All the necessary information on the parameters in the above formula can be 
obtained
from Ref.\cite{UW2}. The eigenvalues of the one-loop anomalous dimensions, 
$\lambda_i^n
(i=+,-,NS)$, are given in Appendix A. The parameters $\tilde{P}_i^n$, 
$A_i^n, B_i^n$ and
$C_\gamma^n$, which are relevant to  the structure function $F_2^\gamma$, 
are given in
Appendix B. Finally the parameters $\tilde{P}_i^{(L),n}$ and 
$B_{\gamma,L}^n$
relevant to the longitudinal structure function $F_L^\gamma$ are given in
Appendix C.

The $n$-th moment of $g_1^\gamma$ up to NLO is expressed in a similar form 
as above is given in Eq.(3.16) of Ref.\cite{SU}.


\newpage


\newpage

\begin{figure}
\centerline{\epsfxsize=8cm
\epsfbox{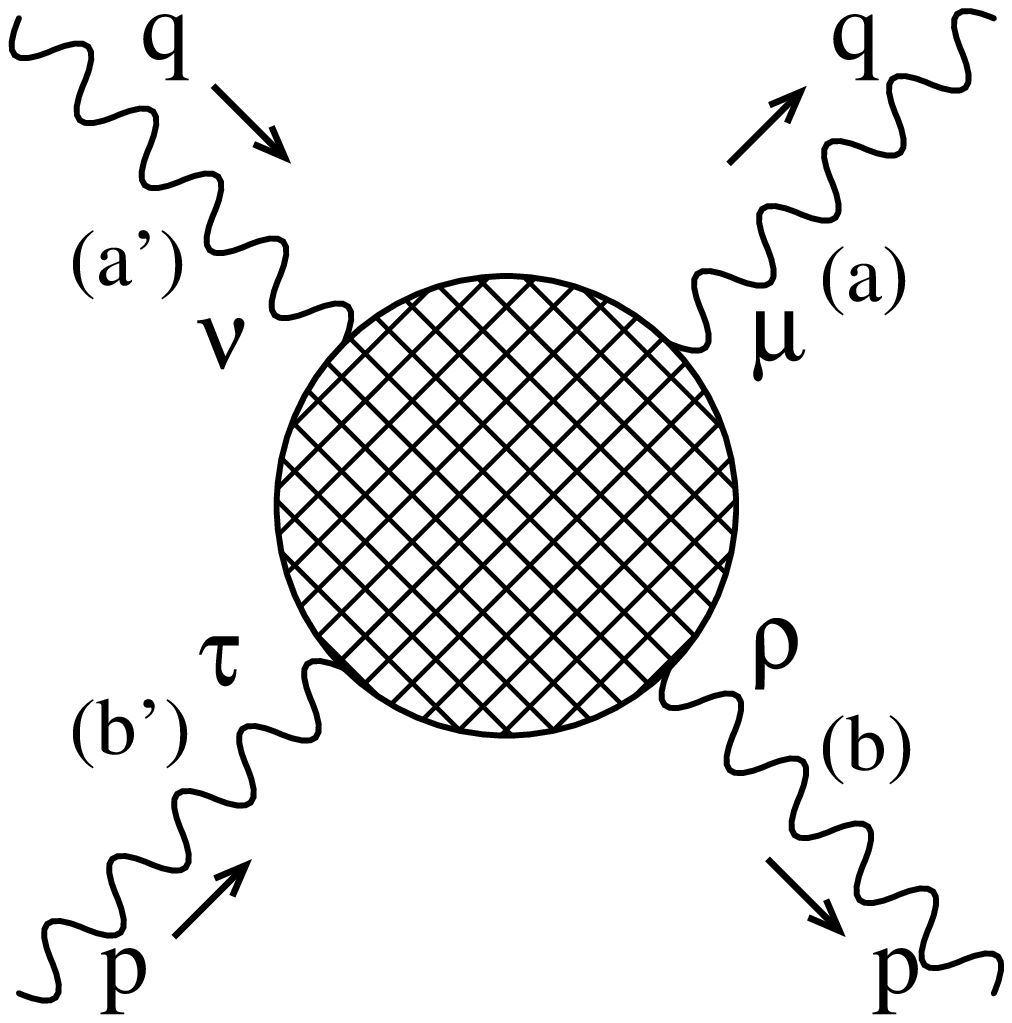}}
\vspace{1cm}
\caption{ Virtual photon-photon forward scattering with momenta $q(p)$ and
helicities $a(b)$ and $a'(b')$.}
\end{figure}

\begin{figure}
\centerline{\epsfxsize=12cm
\epsfbox{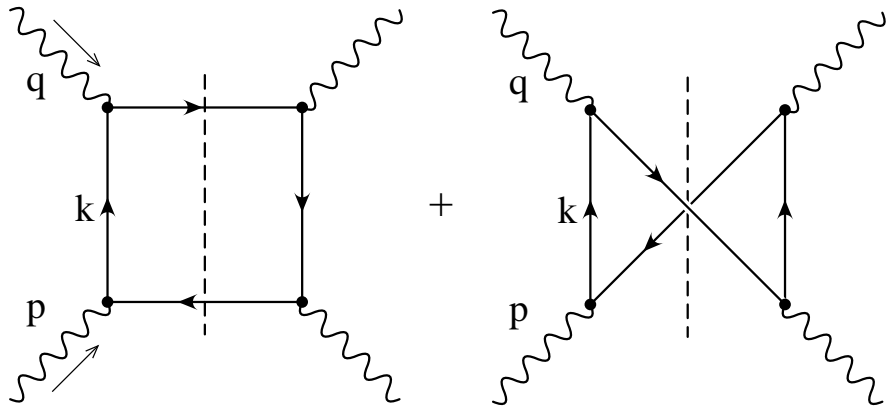}}
\vspace{1cm}
\caption{ The box diagrams in the parton model calculation.}
\end{figure}

\newpage

\begin{figure}
\centerline{\epsfxsize=10cm
\epsfbox{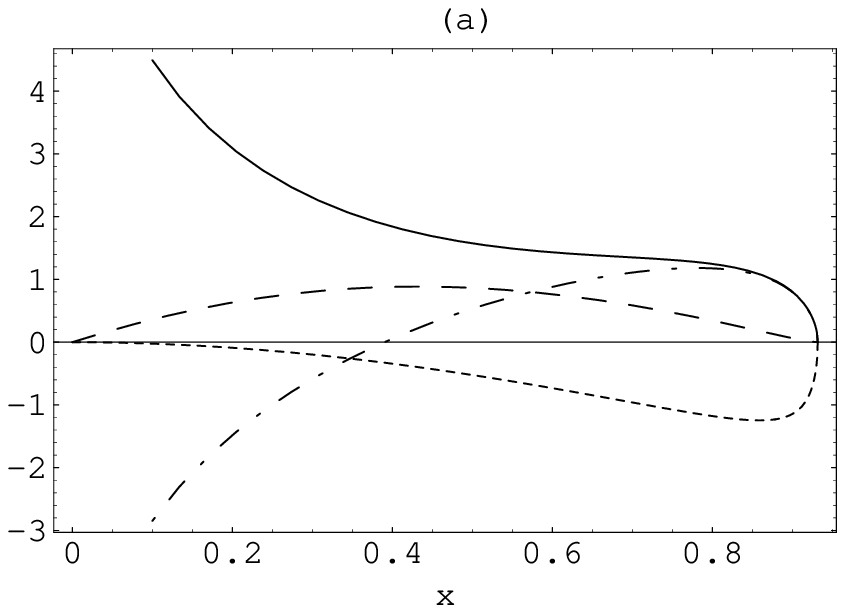}}
\vspace{1cm}
\centerline{\epsfxsize=10cm
\epsfbox{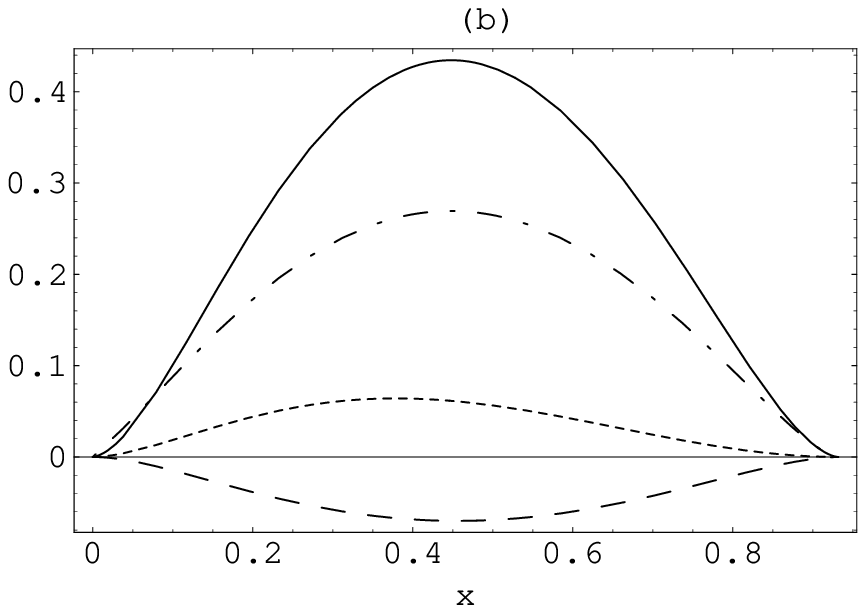}}
\caption{ The PM predictions versus $x$ for the eight virtual photon 
structure functions
in units of  $(\alpha/2\pi)\delta_\gamma$ for $P^2/Q^2=1/30$ and
$m^2/Q^2=1/100$. (a)  $W_{TT}\vert_{PM}$ (solid line),
$W_{TT}^a\vert_{PM}$ (dash-dotted line), $W_{ST}\vert_{PM}$ (long-dashed 
line), and
$W_{TT}^{\tau}\vert_{PM}$ (short-dashed line) ;~
(b) $W_{TS}\vert_{PM}$ (solid line), $W_{TS}^{\tau}\vert_{PM}$
(dash-dotted line), and  $W_{TS}^{\tau a}\vert_{PM}$ (long-dashed line),
and, $W_{SS}\vert_{PM}$ (short-dashed line).
}
\end{figure}

\newpage

\begin{figure}
\centerline{\epsfxsize=10cm
\epsfbox{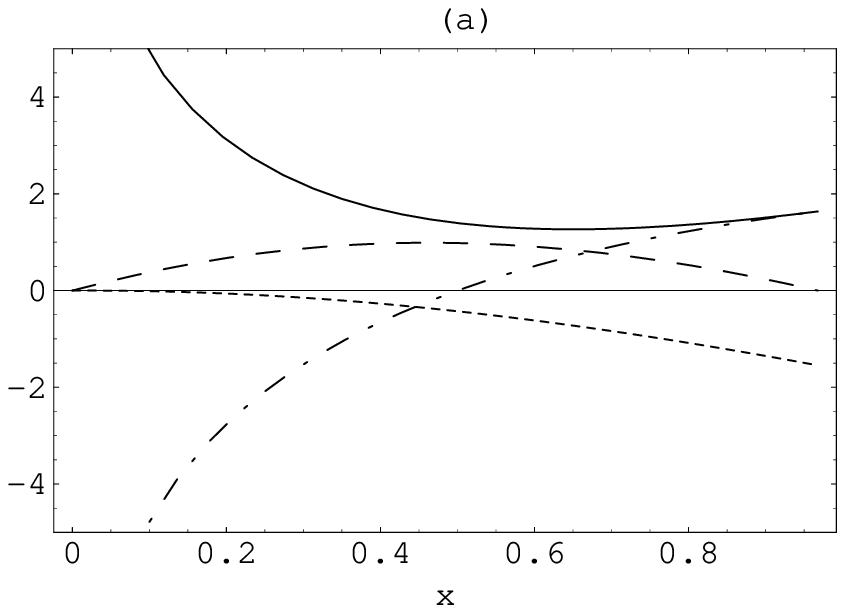}}
\vspace{1cm}
\centerline{\epsfxsize=10cm
\epsfbox{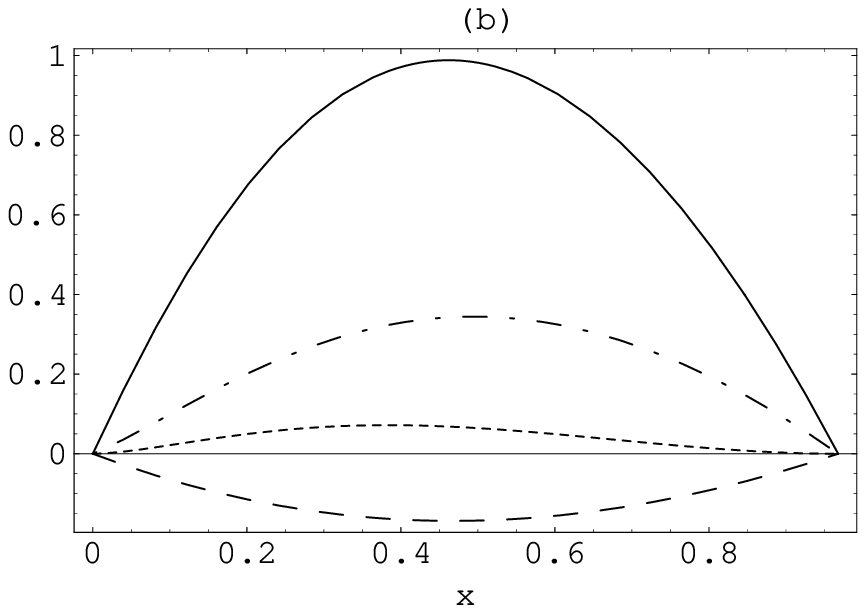}}
\caption{ The PM predictions versus $x$ for the eight virtual photon 
structure functions
in units of  $(\alpha/2\pi)\delta_\gamma$ for massless quark, $m=0$, and 
$P^2/Q^2=1/30$.
(a) $W_{TT}\vert_{PM}$ (solid line),
$W_{TT}^a\vert_{PM}$ (dash-dotted line), $W_{ST}\vert_{PM}$ (long-dashed 
line), and
$W_{TT}^{\tau}\vert_{PM}$ (short-dashed line) ;~
(b) $W_{TS}\vert_{PM}$ (solid line), $W_{TS}^{\tau}\vert_{PM}$
(dash-dotted line), and  $W_{TS}^{\tau a}\vert_{PM}$ (long-dashed line),
and, $W_{SS}\vert_{PM}$ (short-dashed line). Note that 
$W_{ST}\vert_{PM}$ in (a) coincides with $W_{TS}\vert_{PM}$ in (b)
, as they should for $m=0$.
}
\end{figure}

\newpage

\begin{figure}
\centerline{\epsfxsize=9cm
\epsfbox{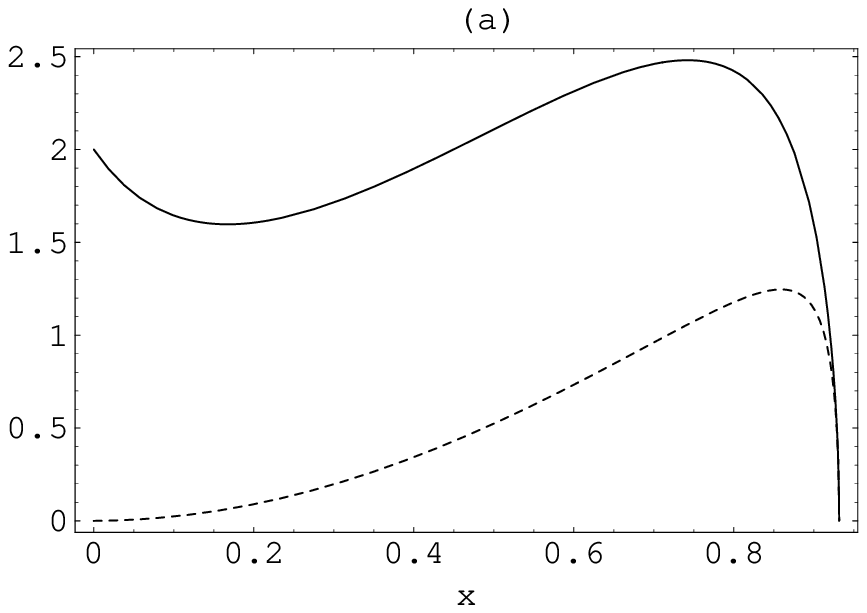}}
\centerline{\epsfxsize=9cm
\epsfbox{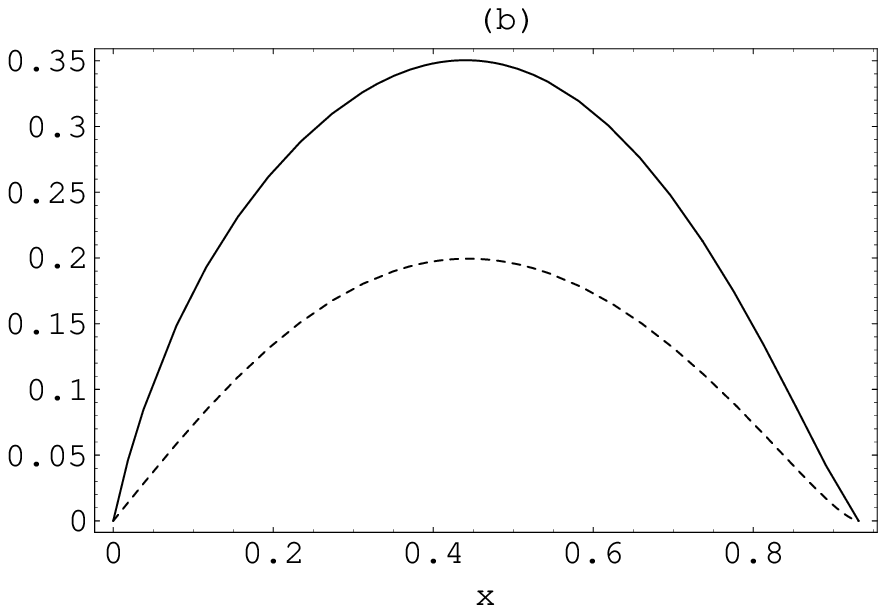}}
\centerline{\epsfxsize=9cm
\epsfbox{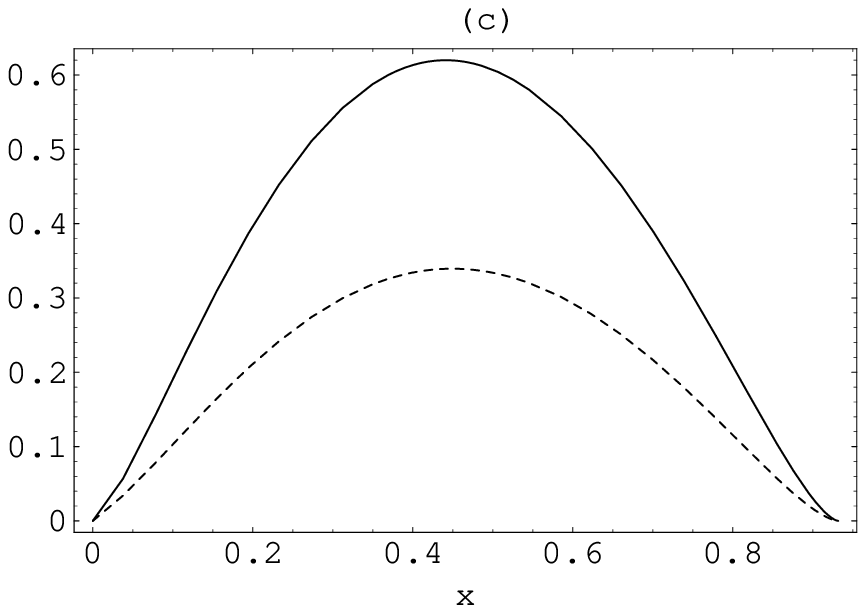}}
\caption{ The positivity constraints and the PM
predictions versus $x$ for $P^2/Q^2=1/30$ and $m^2/Q^2=1/100$.
 The vertical axes are in units of $(\alpha/2\pi)\delta_\gamma$.
(a) $(W_{\rm TT}+W_{\rm TT}^a)$  (solid line) and $\Bigl|W_{\rm TT}^\tau
\Bigr|$ (short-dashed line); (b) $\sqrt{(W_{\rm TT}+W_{\rm TT}^a)W_{\rm SS}}$
   (solid line)  and $\Bigl|W_{\rm TS}^\tau +W_{\rm TS}^{\tau a}  \Bigr|$
(short-dashed line); (c) $\sqrt{W_{\rm TS}W_{\rm ST}}$   (solid line)
and $\Bigl| W_{\rm TS}^\tau-W_{\rm TS}^{\tau a} \Bigr|$ (short-dashed line). }

\end{figure}
\newpage

\begin{figure}
\centerline{\epsfxsize=9cm
\epsfbox{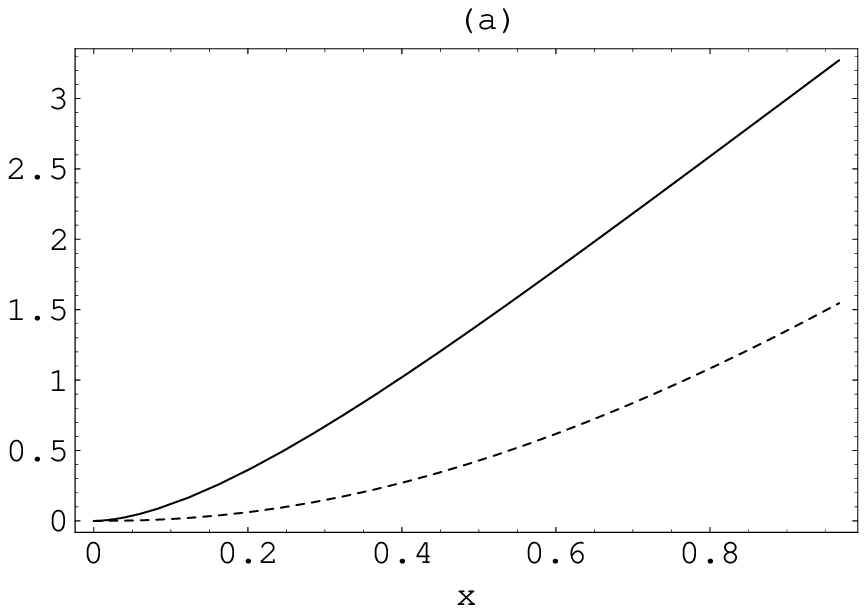}}
\centerline{\epsfxsize=9cm
\epsfbox{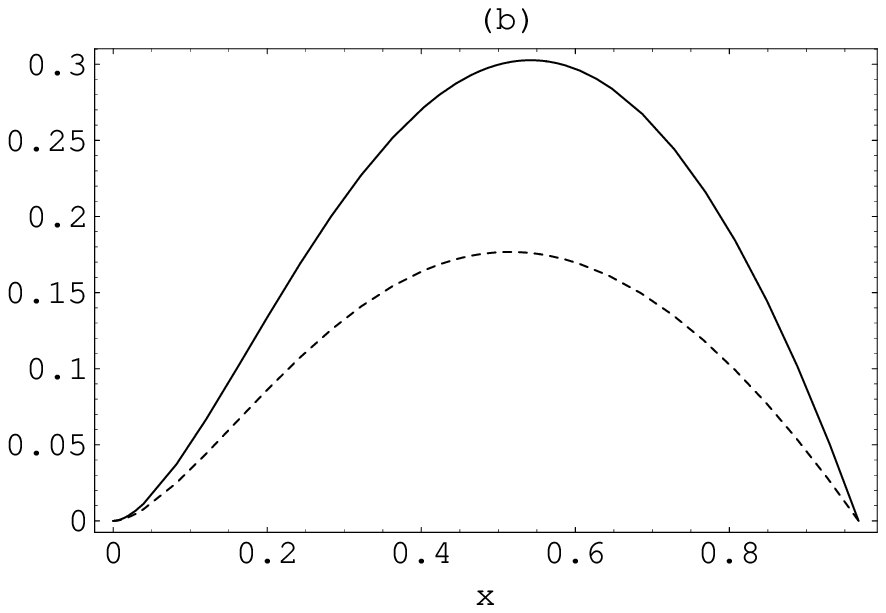}}
\centerline{\epsfxsize=9cm
\epsfbox{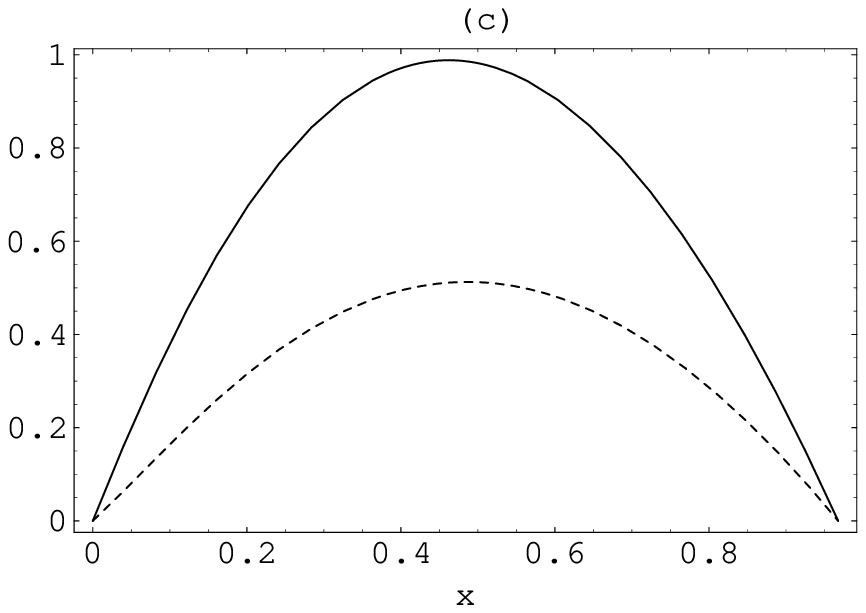}}
\caption{ Same as Fig.5 for massless quark, $m=0$ and $P^2/Q^2=1/30$. 
}

\end{figure}

\begin{figure}
\vspace*{-2cm}
\centerline{\epsfxsize=16cm
\epsfbox{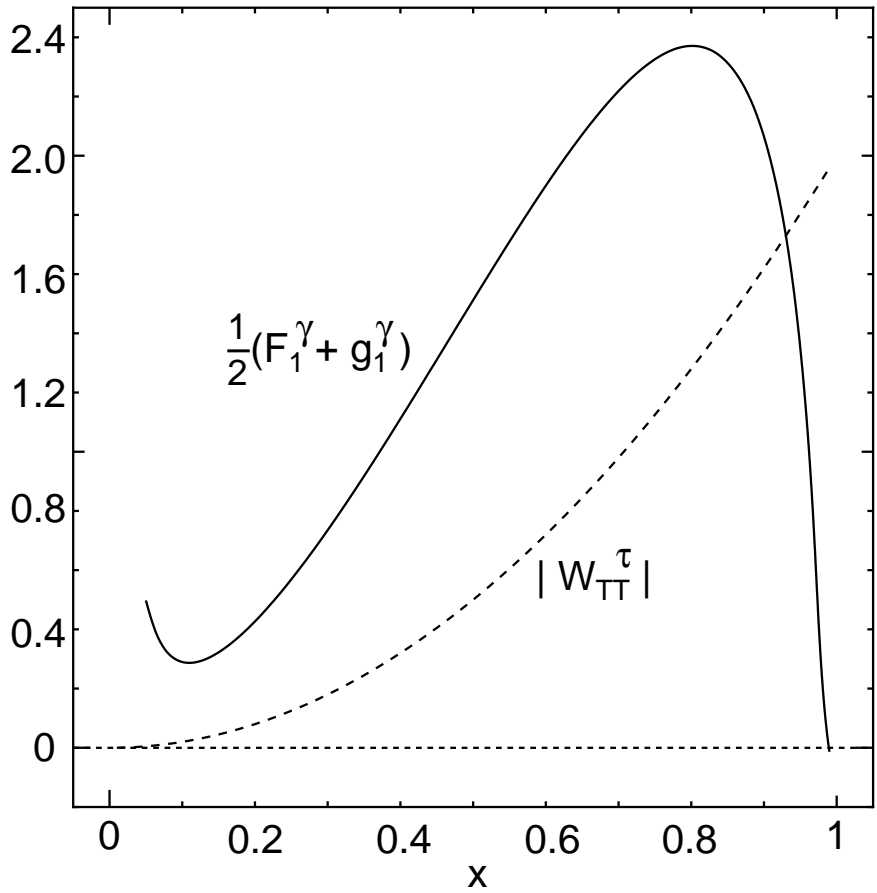}}
\vspace*{-4cm}
\caption{The positivity constraint and the pQCD prediction.
$\frac{1}{2}(F_1^\gamma+g_1^\gamma)$ (solid line) and $\Bigl|W_{\rm TT}^\tau
\Bigr|$ (short-dashed line)
in units of $(\alpha/2\pi)\delta_\gamma$, for $Q^2=30$GeV$^2$,
$P^2=1$GeV$^2$ with $\Lambda=0.2~ {\rm GeV}$ and $N_f=3$.  }

\end{figure}

\end{document}